\documentclass[twocolumn,showpacs,preprintnumbers]{revtex4}
\usepackage{amssymb}

\usepackage[dvips]{graphicx}

\begin{document}

\title{Comment on 'Effect of on-site Coulomb repulsion on superconductivity in the boson-fermion model'
(Phys. Rev. B{\bf 66}, 134512 (2002))}
\author{A.S.~Alexandrov}
\address{Department of Physics, Loughborough University,
Loughborough LE11 3TU, United Kingdom}

\begin{abstract}
The two-dimensional (2D) boson-fermion model (BFM) of
high-temperature superconductors, numerically studied by Domanski
(Phys. Rev. B{\bf 66}, 134512 (2002))
 is not a superconductor.  The critical temperature of the model  is  zero for
any symmetry of the order parameter. The opposite conclusion
advocated by
 Domanski stems from  a  mean-field approximation (MFA) neglecting the boson
 self-energy which is qualitatively erroneous for any-dimensional BFM.
\end{abstract}

 \maketitle

\bigskip

Recently Domaski \cite{dam} and some other authors (see references
in \cite {dam}) claimed that 2D BFM with hybridised fermions and
immobile hard-core bosons is capable to reproduce the phase
diagram of cuprates. The model is defined by the Hamiltonian,
\begin{eqnarray}
H &=&\sum_{{\bf k},\sigma =\uparrow ,\downarrow }\xi _{{\bf k}}c_{{\bf k}%
,\sigma }^{\dagger }c_{{\bf k},\sigma }+E_{0}\sum_{{\bf q}}b_{{\bf q}%
}^{\dagger }b_{{\bf q}}+ \\
&&{\rm g}N^{-1/2}\sum_{{\bf q,k}}\left( \phi _{{\bf k}}b_{{\bf q}}^{\dagger
}c_{-{\bf k}+{\bf q/2},\uparrow }c_{{\bf k}+{\bf q/2},\downarrow
}+H.c.\right) ,  \nonumber
\end{eqnarray}
where $\xi _{{\bf k}}=-2t(\cos k_{x}+\cos k_{y})-\mu $ is the 2D energy
spectrum of fermions, $E_{0}\equiv \Delta _{B}-2\mu $ is the bare boson
energy with respect to the chemical potential $\mu $, ${\rm g}$ is the
magnitude of the anisotropic hybridisation interaction, $\phi _{{\bf k}%
}=\phi _{-{\bf k}}$ is the anisotropy factor, and $N$ is the number of
cells. Ref. \cite{dam} argued that 'superconductivity is induced in this
model from the anisotropic charge exchange interaction (${\rm g}\phi _{{\bf k%
}}$) between the conduction-band fermions and the immobile hard-core
bosons', and 'the on-site Coulomb repulsion $U$ competes with this pairing'
reducing the critical temperature $T_{c}$ less than by 25\%. The author of
Ref. \cite{dam} neglected our study of BFM \cite{ale}, which revealed a
devastating effect of the boson self-energy on $T_{c}$. Here I show that
because of this effect $T_{c}=0K$ in the model, Eq.(1), even in the absence
of the Coulomb repulsion, $U=0$, and the mean-field approximation of Ref.
\cite{dam} is meaningless for any-dimensional BFM.

Using MFA Ref. \cite{dam} decouples bosons and fermions in Eq.(1) replacing
boson operators by $c$-numbers for ${\bf q}=0$. Then $T_{c}$ is numerically
calculated using the linearised BCS equation for the fermionic
order-parameter $\Delta _{{\bf k}}$,
\begin{equation}
\Delta _{{\bf k}}=\frac{{\rm \tilde{g}}^{2}\phi _{{\bf k}}}{E_{0}N}\sum_{%
{\bf k}^{\prime }}\phi _{{\bf k}^{\prime }}{\frac{\Delta _{{\bf k}^{\prime
}}\tanh (\xi _{{\bf k}^{\prime }}/2k_{B}T)}{{2\xi _{{\bf k}^{\prime }}}},%
}
\end{equation}
where $E_{0}$ is determined by the atomic density of bosons ($n^{B}$) \ as
(Eq.(9) in Ref. \cite{dam})
\begin{equation}
\tanh \frac{E_{0}}{2k_{B}T}=1-2n^{B},
\end{equation}
with $T=T_c$ and ${\rm \tilde{g}}^{2}={\rm g}^{2}(1-2n^{B})$.
While Eq.(2) is perfectly correct, Eq.(3) is incorrect because the
boson self-energy $\Sigma _{b}({\bf q},\Omega _{n})$ due to the
exchange interaction is missing. At first sight \cite{dam} the
self-energy is small if ${\rm g}$ is small in comparison to the
kinetic energy of fermions. However $\Sigma _{b}(0,0)$ diverges
logarithmically at zero temperature \cite{ale}, no matter how week
the interaction is. Therefore it should be kept in the density
sum-rule, Eq.(3). Introducing the boson Green's function
\begin{equation}
D({\bf q},\Omega _{n})=\frac{1-2n^{B}}{i\Omega _{n}-E_{0}-\Sigma _{b}({\bf q}%
,\Omega _{n})}
\end{equation}
one must replace incorrect Eq.(3) by

\bigskip
\begin{equation}
-{\frac{k_{B}T}{{N}}}\sum_{{\bf q},n}e^{i\Omega _{n}\tau }D({\bf q}%
,\Omega _{n})=n^{B},
\end{equation}
where $\tau =+0$, and $\Omega _{n}=2\pi k_{B}Tn$ ($n=0,\pm 1,\pm
2...$). The divergent (cooperon) contribution to $\Sigma _{b}({\bf
q},\Omega _{n})$ is given by \cite{ale}
\begin{eqnarray}
&&\Sigma _{b}({\bf q},\Omega _{n})=-\frac{{\rm \tilde{g}}^{2}}{2N}\sum_{{\bf %
k}}\phi _{{\bf k}}^{2}\times  \\
&&\frac{\tanh [\xi _{{\bf k-q/2}}/(2k_{B}T)]+\tanh [\xi _{{\bf k+q/2}%
}/(2k_{B}T)]}{\xi _{{\bf k-q/2}}+\xi _{{\bf k+q/2}}-i\Omega _{n}},
\nonumber
\end{eqnarray}
so that one obtains
\begin{equation}
\Sigma _{b}({\bf q},0)=\Sigma _{b}(0,0)+\frac{q^{2}}{2M^{\ast }}+{\cal O}%
(q^{4})
\end{equation}
for small ${\bf q}$ with any anisotropy factor compatible with the
point-group symmetry of the cuprates. Here $M^{\ast }$ is the boson mass,
calculated analytically in Ref.\cite{ale} with the isotropic exchange
interaction and parabolic fermion band dispersion (see also Ref.\cite{cris}%
), and $\hbar=1$. The BCS-like equation (2) has a nontrivial
solution for $\Delta _{{\bf k}} $, if
\begin{equation}
E_{0}=-\Sigma _{b}(0,0).
\end{equation}
Substituting Eq.(7) and Eq.(8) into the sum-rule, Eq.(5) one obtains a
logarithmically divergent integral with respect to ${\bf q}$, and
\begin{equation}
T_{c}=\frac{const}{\int_{0}dq/q}=0.
\end{equation}
Using a 'bubble' approximation for the self-energy Ref.\cite{ale}
proved that the Cooper pairing of fermions in BFM is impossible
without the Bose-Einstein condensation (BEC) of real bosons. The
bubble approximation is actually exact because of the logarithmic
divergence of the Cooperon diagram, as  was also confirmed in
Ref.\cite{ren}. Hence, the devastating result, Eq.(9) is a direct
consequence of the well-known theorem, which states that BEC is
impossible in 2D.

One may erroneously believe that MFA results \cite{dam} can be
still applied in three-dimensions, where BEC is possible. However,
increasing dimensionality does not make MFA a meaningful
approximation for the boson-fermion model. This approximation
leads to a naive conclusion that a BCS-like superconducting state
 occurs below the
 critical temperature   $T_{c}\simeq \mu \exp\left( -{%
E_{0}/z_c}\right) $ via fermion pairs being \emph{virtually}
excited into
 $unoccupied$ bosonic states \cite{lee,ran}.  Here $z_c=\tilde{g}^{2}N(0)$ and
$N(0)$ is the density of states (DOS) in the fermionic band near
the Fermi level $\mu $. However,  the Cooper pairing of fermions
 is impossible via virtual unoccupied bosonic states. It occurs
only simultaneously with the Bose-Einstein condensation of real
bosons in the exact theory of 3D BFM \cite{ale}.

The origin of the simultaneous condensation of the fermionic and
bosonic fields in 3D BFM lies in the  softening of the boson mode
at $T=T_c$ caused by its hybridization with fermions. Indeed,
Eq.(8) does not depend on the dimensionality, so that the
analytical continuation of Eq.(4) to real frequencies $\omega$
yields the partial boson DOS as $\rho(\omega)=(1-2n_B)
\delta(\omega)$ at $T=T_c$ for ${\bf q}=0$ in any-dimensional BFM
for any coupling with fermions.

Taking into account the boson damping and dispersion shows that
the boson spectrum significantly changes for all momenta.
Continuing the self-energy, Eq.(6) to real frequencies yields  the
damping (i.e. the imaginary part of the self-energy) as \cite{ale}
\begin{equation}
\gamma({\bf q},\omega)={\pi z_c\over{4q\xi}} \ln
\left[{\cosh(q\xi+\omega/(4k_{B}T_{c}))\over{\cosh(-q\xi+\omega/(4k_{B}T_{c}))}}\right],
\end{equation}
where $\xi=v_F/(4k_{B}T_{c})$ is a coherence length. The damping
is significant when $q\xi<<1$. In this region $\gamma({\bf
q},\omega)=\omega\pi z_{c}/(8k_{B}T_{c})$ is comparable or even
larger than  the boson energy $\omega$. Hence bosons look like
overdamped diffusive modes, rather than quasiparticles in the
long-wave limit \cite{ale,cris}, contrary to the erroneous
conclusion of Ref.\cite{ran0}, that there is 'the onset of
coherent free-particle-like motion of the bosons' in this limit.
Only outside the long-wave   region, the damping becomes small.
Indeed, using Eq.(10) one obtains $\gamma({\bf q},\omega)=\omega
\pi z_{c}/(2qv_F)<< \omega$, so that bosons at
 $q >>1/\xi$ are well defined quasiparticles
 with a logarithmic dispersion, $\omega(q)=z_c \ln(q
\xi)$ \cite{ale}.  As a result the boson dispersion is distributed
over the whole energy interval from zero up to $E_0$,
 but not a delta-function at $E_0$ even in the weak-coupling limit.

The main mathematical problem with MFA stems from the density sum
rule, Eq.(5) which determines the chemical potential of the system
and consequently the bare boson energy $E_{0}(T)$ as a function of
temperature. In the framework of MFA one takes the bare boson
energy   in Eq.(2) as a temperature independent parameter,
$E_0=z_c\ln (\mu/T_c)$, or determines it from the conservation of
the total number of particles, Eq.(5) neglecting the boson
self-energy $\Sigma_b({\bf q}, \Omega)$ \cite{lee,ran,dam}). Then
Eq.(2) looks like the conventional mean-field BCS equation, or the
Ginzburg-Landau equation (near the transition) with a negative
coefficient $\alpha \propto T-T_c$ at $T<T_c$ in the linear term
with respect to $\Delta(T)$. Hence, one concludes that the phase
transition is almost the conventional BCS-like transition, at
least at $E_0\gg T_c$ \cite{lee,ran}, and, using the Gor'kov
expansion in powers of $\Delta$, finds  a finite upper critical
field $H_{c2}(T)$  \cite{dam2}. These findings are mathematically
and physically wrong. Indeed, the term of the sum in Eq.(5) with
$\Omega_n=0$ is given by the integral
\begin{equation}
T\int {d{\bf q}\over{2\pi^3}}{1\over{E_0+\Sigma_b({\bf q},0)}}.
\end{equation}
 The integral converges, if and
only if $E_0\geqslant -\Sigma_b(0,0)$. In fact,
$E_0+\Sigma_b(0,0)$ is strictly zero in the Bose-condensed state,
because $\mu_b=-[E_0+\Sigma_b(0,0)]$ corresponds to the boson
chemical potential relative to the lower edge of the boson energy
spectrum. More generally, $\mu_b=0$ corresponds to the appearance
of the Goldstone-Bogoliubov mode due to a broken symmetry  below
$T_c$. This exact result makes the BSC equation (2) simply an
identity \cite{ale} with  $\alpha \equiv 0$ at any temperature
below $T_c$. On the other hand, MFA
 violates the density
sum-rule, predicting the wrong negative $\alpha(T)$ below $T_c$.

 Since $\alpha(T)=0$,
 the Levanyuk-Ginzburg parameter
 \cite{lev} is infinite, $Gi=\infty$. It means that the  phase
 transition is never a BCS-like second-order phase transition
 even at large $E_0$ and small $g$. In fact, the
 transition  is driven by the Bose-Einstein condensation of \emph{
 real} bosons with ${\bf q}=0$, which occur  due to the complete
 softening of their spectrum at  $T_c$ in 3D BFM.
 Remarkably, the conventional upper critical field, determined as the field, where a non-trivial
 solution of the linearised Gor'kov equation   occurs, is
 zero in BFM, $H_{c2}(T)=0$, because $\alpha(T)=0$ below $T_c$. It is not  a finite $H_{c2}(T)$
 found in Ref. \cite{dam2} using MFA.  Even
 at  temperatures well below $T_c$ the condensed state is fundamentally
 different from the BCS-like MFA ground state, because of the \emph{pairing} of
 bosons \cite{alecon}.    The pair-boson condensate
 significantly
modifies the thermodynamic
  properties of the condensed BFM
  compared with the MFA predictions.

This qualitative failure of MFA  might be rather unexpected, if
one believes that bosons in Eq.(1) play the same role as phonons
in the BCS superconductor. This is not the case for two reasons.
The first one is  the density sum-rule, Eq.(5), for bosons which
is not applied to phonons. The second being that the boson
self-energy is given by the divergent (at $T=0$) Cooperon diagram,
while the self-energy of phonons is finite at small coupling.

I have to conclude that the numerical work by Domanski \cite{dam}
does not make any sense in any dimension. There is nothing in 2D
BFM to compete with because the model is not a superconductor even
without the Coulomb repulsion. The MFA results \cite{dam,dam2} do
not make any sense in three dimensions either, because the
divergent self-energy has been neglected in calculating $T_{c}$
and $H_{c2}(T)$. The common wisdom that at weak coupling  the
boson-fermion model is adequately described by the BCS
 theory, is  negated by our results.

\end{document}